\documentclass[12pt]{article}
\usepackage{rotating}
\usepackage{epsfig,amssymb}
\usepackage{ulem}
\usepackage{amsmath}
\input{epsf}

\def\laq{\raise 0.4 ex \hbox{$<$}\kern -0.8 em\lower 0.62 ex\hbox{$\sim$}}
\def\gaq{\raise 0.4 ex \hbox{$>$}\kern -0.7 em\lower 0.62 ex\hbox{$\sim$}}

\def\beq{\begin{equation}}
\def\eeq{\end{equation}}
\def\beqa{\begin{eqnarray}}
\def\eeqa{\end{eqnarray}}

 \def\frac#1#2{{\textstyle{{#1}\over {#2}}}}
 \def\lsim{\mathrel{\rlap{\lower4pt\hbox{\hskip1pt$\sim$}}
    \raise1pt\hbox{$<$}}} \def\gsim{\mathrel{\rlap{\lower4pt\hbox{\hskip1pt$\sim$}}
    \raise1pt\hbox{$>$}}}
\def\sqr#1#2{{\vcenter{\vbox{\hrule height.#2pt
         \hbox{\vrule width.#2pt height#1pt \kern#1pt
         \vrule width.#2pt}
         \hrule height.#2pt}}}}

\def\gappeq{\mathrel{\rlap {\raise.5ex\hbox{$>$}} {\lower.5ex\hbox{$\sim$}}}}

\def\lappeq{\mathrel{\rlap{\raise.5ex\hbox{$<$}}
{\lower.5ex\hbox{$\sim$}}}}

\usepackage{color}

\begin{document}
\pagestyle{plain}

\begin{flushright}

\end{flushright}
\vspace{15mm}

\begin{center}

{\Large\bf On Bayesian inference considerations and other issues concerning Drake's equation of  Astrosociobiology}

\vspace*{1.0cm}

Orfeu Bertolami$^{1,2}$ \\
\vspace*{0.5cm}
{$^{1}$ Departamento de F\'{\i}sica e Astronomia, Faculdade de Ci\^encias,
Universidade do Porto, \\
Rua do Campo Alegre s/n, 4169-007 Porto, Portugal}\\

{$^{2}$ Centro de F\'{\i}sica das Universidades do  Minho e do Porto,
Rua do Campo Alegre s/n, 4169-007 Porto, Portugal}\\

\vspace*{1.0cm}
\end{center}

\begin{abstract}
\noindent
Speculation about the existence of  advanced forms of life in the Universe and  in our galaxy, has been since ever a subject of fascination and  discussion in fiction, as well as in astrophysics, biology and philosophy. The well-known Fermi's 1950s challenge, "Where are the aliens?" has acquired more substance with the realisation of the potentialities of radioastronomy, which led to the paradigmatic Drake's equation. The emergence of astrobiology,  together with the discovery up to now of more than seven thousand exoplanets, has brought increasing support to the discussion about putative life cradles. However, after more than six decades, the only quantitative tool available to estimate how widespread is life and, in particular, advanced forms of life, is, besides direct searches, which so far provided no evidence,  still Drake's equation. In the present work we review the current knowledge about this equation and present new arguments of multiple origin in order to evaluate one of its most critical terms, namely the one associated to the time span that a technological civilisation must search for detectable signs of the existence and for how long a search must be extended to bear fruits. We propose that this term should be replaced by a more specific one which involves critical parameters in the enterprise of gathering information, such as  energy expenditure, searching area and entropy generation. These terms can be regarded as the capability that any cosmic civilisation must show in order to face the challenge of going beyond the climate and other crises that its development inevitably ensues. 
Our considerations suggest that a typical time span is about a couple of decades, meaning  that a successful and systematic searching programme around about hundred stars might take around a few thousand years. 
\end{abstract}

\vfill
\noindent\underline{\hskip 140pt}\\[4pt]
\noindent
{E-mail address: orfeu.bertolami@fc.up.pt}

\newpage

\section{Introduction}
\label{sec:introduction}

Humankind was awaking from the nightmare of two world wars and already facing the possibility of a direct nuclear confrontation between EUA and Soviet Union, when the scientific community realised that radioastronomy could potentially acquire empirical evidence about the centuries old discussion about the existence of life in the Universe (see, for instance, Ref. \cite{BUniverse} for a general and historical discussion). In science, the antagonism West-East, as it was often pictured, existed in theoretical physics, in mathematics and it was most visible due to an open and highly political space race. It was in this context  that a paper authored by Giuseppe Cocconi and Philip Morrison with the provocative title ``Searching for Interstellar Communications" appeared in 1959 \cite{CM} in the Nature journal. The authors argued that radio telescopes were already sensitive enough to capture transmissions that might be broadcast into space by civilisations orbiting other stars. These messages, they suggested, should more likely to be transmitted at a wavelength of 21 cm (1,420.4 MHz), the wavelength of radio emission by neutral hydrogen, the most common element in the Universe, a logical choice for any civilisation seeking cosmic contact. In fact, philosophers, astronomers and physicists have speculated throughout the centuries about the existence and the number of planets in the Universe, and indeed, just a few months later an effective search was carried out, from April to June of 1960, by the radio-astronomer Frank Drake using the 26 m dish of the National Radio Astronomy Observatory, Green Bank, in West Virginia. The search consisted of monitoring two nearby Sun-like stars, Epsilon Eridani and Tau Ceti, slowly scanning frequencies close to the 21 cm wavelength for six hours per day. Subsequently,  in 1961 the first conference on the search for extraterrestrial intelligence was held at Green Bank and it was in this context that the equation that bears Drake's name arose out of his planing for the meeting \cite{D}:
\begin{equation}
N = R_S f_P n_e f_L f_I f_T L~,
\label{eq:I}
\end{equation}
where $N$ is the number of putative civilisations in the Milky Way with which communication might be possible, $R_S$ is the rate of star formation in the Milky Way, $f_P$ is the fraction of stars with planetary systems, $n_e$ is the number of planets that can potentially support life, $f_L$ denotes the fraction of planets that could support life that actually develop life, $f_I$ is the fraction of planets that develop civilised forms of intelligent life, $f_T$ is the fraction of civilisations that develop technology that allows them to emit detectable signs of the their existence in the cosmos, and $L$ the time interval for which such civilisations emit detectable signs into space. This speculative, but simple equation can yield a wide range of results. The discovery of exoplanets and the knowledge about their formation and development may shed some light in some terms of the Drake equation. A slightly different version of this equation was proposed by Carl Sagan \cite{S}, by replacing $R_S$ directly by the number of stars in our galaxy, $N_S$. Another alternative was proposed in Ref. \cite{Loeb}. 
In what follows we shall also propose a counterpoint to the Drake equation, more particularly in what concerns its last term and we shall argue that many of the remaining terms can be simplified through Bayesian arguments. 

Before discussing estimates for $N$, it is relevant to spell out some of the assumptions that support the objectivity of any Drake-type equation. The first point to consider is the universality of the laws of physics and hence of chemistry. This universality or covariance, if one prefers, is supported by well established symmetries such as the weak equivalence principle, local Lorentz invariance, the  and local position invariance (see, for instance, Ref. \cite{OBP} for a discussion). This universality ensures that, like any human, any sentient member of a cosmic ``community" will estimate the physical parameters of Drake's equation in the same way and reach similar conclusions in what refers to their values. Conclusions concerning biological, geological, anthropological, historical and sociological type parameters are not necessarily as universal, being, by their very nature, more contingent on 
genetic dependence of the multiple development processes involved. That is why Drake's equation is somewhat more than just a physical, astrophysical or astrobiology equation. It also contains anthropological, biological, geological and socio-historical factors being hence somewhat more than just a tool for astrobiology. and should be regarded as an astrosociobiology equation.  Nevertheless, despite of that it will be assumed, under somewhat well defined conditions, that any generic member of a cosmic ``community" will most likely use a close version of the Drake equation in order to estimate the probability of establishing communication with other members of the civilisations that putatively might exist out there. Reversibility of the electromagnetic radiation imply that in or out signals share the same features.

As one could already anticipate, estimates on the value of $N$ admit a wide variation. Educated assessment of the various terms yields $20 \lsim  N  \lsim 5 \times 10^7$ (see Ref. \cite{Schilling} and refs. therein). Estimates of the factors of the Drake equation based on Monte Carlo simulations of estimates of a stellar and planetary model of the Milky Way yield that the number of civilisations varies by a factor of 100 \cite{F}.  The well-known Fermi's paradox about the lack of evidence on the existence of extraterrestrial civilisations, suggests that in our galaxy $N=1$. Crucial arguments in favour of the concept that life is a rare event in the Universe suggests an universal result $N<<1$ and $N=1$ in our galaxy, or otherwise, assuming that life is a cosmic imperative, as we defended elsewhere \cite{OB2006}, would mean that $N>>1$, the Giordano Bruno solution \cite{Bruno}. But, of course, it should be understood that Drake's equation, as already pointed out, has an hybrid character as it contais terms related with the conditions for the onset of life in the Universe as well as terms directly related to the technological development of civilisations, while the last term concerns the willingness of technically advanced civilisations to search and establish contact with other cosmic civilisations. This means that in order to estimate the output of Drake's equation one has to consider astrophysical, biological, anthropological, historical and sociological as well as technical aspects of communication in the radio bandwiidth.  

After Drake's equation, the discussion was further boosted when, in 1964, Soviet astrophysicist Nikolai Kardashev \cite{K}, proposed, in a review of the Soviet radioastronomy, a scale to gauge the degree of scientific and technological achievement of a cosmic civilisation, at the Byurakan conference in Armenia.  His paper adopts human civilisation as a model to be extrapolated and the immutability and universality of physical laws to propose the following classification, nowadays known as the Kardashev classification:  

\noindent
A Type I civilisation is one that is capable to access and store all the energy available on its planet and also to control the natural events such as earthquakes, volcanic eruptions, etc. This has a indelible connection with the Anthropocene age we have entered in 1950s \cite{Crutzen}; 

\noindent
A Type II civilisation is one that can actively handle a significant fraction of the energy from its star, through the use, for instance, of a Dyson sphere \cite{Dyson} and other technological means; 

\noindent
A Type III civilisation, is one that is able to capture and manipulate energy equivalent to all components of a galaxy, that is of every star, black hole, etc, within it. 

A more quantitative version of the Kardashev classification has been proposed by Carl Sagan (see, for instance Ref. \cite{Carrigan}), according to which the stage of development of a civilisation could be evaluated through the logarithmic interpolation formula: 
\begin{equation}
K = {\log E_T -6 \over 10} ~, 
\label{eq:K}
\end{equation}
where $E_T$ is the overall power output of the planet. In this more quantitative evaluation, Earth is not yet a Type I civilisation, but  a $K=0.72$ civilisation.

Subsequently, Kardashev has considered scenarios for investigating hypothetical extraterrestrial super civilisations \cite{K1}. 

On the other hand, a more recent discussion on Drake's equation consider the progress on the astrophysical terms of the equation and on the search of exoplanets to argue that the human civilisation is likely to be unique in the cosmos if the odds of a civilisation developing on a habitable planet are less than $10^{-22}$ \cite{Frank1}. 
In fact, it has been pointed out that from the biological point of view the crux of matter is determined by the rate of increase of genetic complexity \cite{Sharov_Gordon}. If this rate is slow, then life on Earth is a singular cosmic event, that is, $N = 1$, depriving the very meaning of Drake's equation; if otherwise, the  rate of increase of genetic complexity is high and hence life is cosmically abundant, many observational and theoretical avenues are possible. Recent announcement that markers of biological activity have been detected in the K2-18 b exoplanet might indicate that that genetic complexity is fast growing, which also opens the very exciting possibility of observing life markers at the edges of the transit motion of exoplanets around its star as observed by the James Webb telescope \cite{Madhusudhan}. In fact. the various observational perspectives were reviewed in the Astro 2020 White Papers of the National Academies of Science, Engineering and Medicine \cite{Astro_2020}. For a critical assessment on the research disparity of funding for astrobiology and its public perception, see e.g. Ref. \cite{Eldadi}. Of course, programmes for the Search for Extraterrestrial Intelligence (SETI) has nowadays greatly evolved in what concerns  its capability to detect narrowband radio signals with "techno-signatures" through the use of powerful radio telescopes and arrays of them and, of course, in handling vast amounts of data and in analysing algorithms. The possibility of using powerful laser signals it is also a matter of discussion in SETI research. 

Getting back to the humankind's context in which Drake's equation has emerged, it is just logic to consider the equivalence with the current Anthropocene and the likelihood that the Anthropocene is a transition from the Holocene to a Hot-Earth scenario. Indeed, given the current situation of the climate crisis, it is just natural that the climate anthropogenic risks that emerged with the Anthropocene have replaced the threat of a nuclear crisis of past discussions of Astrobiology\footnote{Even though the recent geopolitical tension created by the criminal invasion of Ukraine by Russia and the Middle East conflict might lead us again to the possibility of a nuclear confrontation.}. Nowadays, the discussion must necessarily consider strategies of evading the risks unleashed by the Anthropocene such as environmental catastrophic events due to the climate crisis, pandemics, socio-economic crises generated by mass unemployment caused by the widespread use of artificial intelligence and robotics. These threats have been referred to as the "great filter" \cite{Jiang} (see also Refs. \cite{Frank2,Jacob}). Clearly, the Anthropocene bottleneck and the danger it represents in terms of the lack of a sustainable future is much more acute than those associated, for instance, to the relatively long term genetic degradation of species and, in particular, of humans. Recent indication of a possible chaotic behaviour of the Earth System due to a hyper human activity \cite{Bernardini} suggests that a long term survival of humankind might be under risk. In fact, it has also been recently argued that once a technological society reaches the Anthropocene, it cannot overcome the limitations of the Second Law of Thermodynamics and will inevitably undermine the very conditions of its habitability, as thoroughly discussed in Ref. \cite{Balbi}. We shall get back to this question in Section 4. In any case, even though the risks posed by the Anthropocene are very palpable, they are not necessarily inevitable. Scientific and technological development might very well get materialised along different and more sustainable paths. 

Thus, after the Cosmological Anthropic Principle \cite{Barrow} that considered the existence of observers, us, as a condition for explaining the features of the observable Universe and the implications for very setting up of the laws of physics, the Anthropocene has put us face to face with the dire consequences for the habitability of the planet due to a voracious technological society driven by consumption, and abundant and cheap fossil fuels. One could argue in favour of a cosmic responsibility \cite{OB10,OB18} in order to preserve the Earth System for all species, but that is meaningless till it can materialise into actions and well defined stewardship planetary goals.

In fiction, it was recently suggested, in the second novel, {\it Dark Forest} \cite{Dark} of the trilogy {\it The Three Body Problem}, that the answer to the Fermi's question is that an enlightened cosmic civilisation understands the risk of revealing its precise location so to avoid the likelihood of being enslaved or destroyed by more advanced civilisations, a reasoning that appeared in a previous science fiction novel {\it The Killing Star} by Charles R. Pellegrino and George Zebrowski \cite{KStar}. The idea of a sociology of civilisations, ruled by conservation laws \cite{Dark} does have, as we shall see below, clear implications for the statistics of stars or planets in equilibrium and controlled by a common environmental variable.  

In this work we shall pursue the idea of estimating the number N, using a suitably modified version of the Drake equation to evaluate the probability of coming across the manifestation of existence of a cosmic civilisation. In order to perform that we shall consider the last term of Drake's equation separately and argue that the others six terms can be suitably estimated on the basis of stellar formation arguments and Bayesian inference  considerations. As we shall see this will naturally provide us with a lower bound on the time necessary to gather evidence about the existence of a cosmic civilisations capable to communicate. Our main conclusion is that, in our galaxy, a Type I Kardashev civilisation willing to spend a reasonable amount of energy for the task of broadcasting-searching the presence of listeners must dedicate itself for the task continuously and for at least a couple of decades in oder to achieve fruitful results.

\section{The Probability Distribution subjected to Conservation Laws in Equilibrium}
\label{sec:iConservation}

For sure any estimate about $N$ crucially depends on the assumptions under consideration. Indeed, let us exemplify this point considering first the distribution functions, that is the statistics of occurrence of outcomes in a set of objects subjected  to conservation laws. If the former can be characterised by a continuous physical variable, say energy, for the case of a set of stars or planets in equilibrium under some environmental physical conditions, most often the temperature of the environment, then it follows that the number of classical objects in the energy range from $E$ to $E+ dE$ is given by the Boltzmann distribution: 
\begin{equation}
n(E) = n_0 \exp[-E/E_0]~,
\label{eq:Boltzmann}
\end{equation}
where $n_0$ corresponds to the density of objects per unit of volume and $E_0$ to a reference value that indicates equilibrium. In statistical physics situations, $E_0=k_B T$, where, $k_B$ corresponds to the Boltzmann's constant and $T$ the temperature. As is well-known, this distribution for classical (non-quantum) distinguishable objects arise out the hypotheses that the number of objects is conserved and their individual energy adds up to a fixed value.  

It would be interesting if in evaluating the various terms in the Drake equation the underlying physics of the problem of setting up conditions for the development and evolution of technologically advanced civilisations, one  could rely on a powerful tool such as the Boltzmann statistics. However, as far as we can imagine, given the lack of interaction and any obvious equilibrium conditions among the civilisations that might exist in the Galaxy, we cannot consider equilibrium arguments in order to evaluate the terms of the Drake's equation.

However, as we shall see, an important simplification in the analysis arise from the fact that in Drake's equations all terms, but the first and the last ones, are determined by conditions that must necessarily be previously satisfied. Hence, in what follows we consider the terms in question of Drake's equation that are subjected to Bayesian inference arguments.

\section{Bayesian considerations}
\label{sec:iEstimateI}

In order to consider a realistic estimate of the factors in Drake's equation, let us start identifying the terms that are directly related with probabilities. It is easy to see that all terms but the first and the last ones can be expressed in terms of probabilities, in fact conditional probabilities. 

The first term in Drake's equation concerns the rate of star formation, for instance, in our Galaxy. Evidence shows that the star formation rate is proportional to the density of the molecular gas, predominantly, hydrogen gas. The Milky Way has a somewhat low rate of star formation. The details are complex and do not allow for a clear cut estimate. In principle, the relevant quantity to consider is the initial mass function, in fact the probability density function of stars of mass, $m$, described by the stars initial mass distribution, $\xi_S(m) dm$, given by the Salpeter function \cite{Salpeter}, which is expressed by a power-law:    
\begin{equation}
\xi_S(m) dm = \xi_{S0} \left( {m \over M_{\odot}}\right)^{-\alpha} {dm \over M_{\odot}}~,
\label{eq:Salpeter}
\end{equation}
where $\xi_{S0}$ is a constant, $M_{\odot}$ is the solar mass and $\alpha \approx 2.35$. Even though this function is not obviously invariant and may not be normalisable, it is assumed that after a suitable integration, which might not be free of difficulties, it yields a figure of merit, say $N_T$ times a power $({m/M_{\odot}})^{-\alpha+1}$, where $N_T$ is suitable normalisation factor (see e.g. Ref. \cite{Kroupa} for a more recent discussion). However, to avoid these complexities, we shall consider as suggested by Sagan \cite{S}, that G-type stars being spectroscopically similar to our sun are the ones in the Main Sequence that provide better changes of success in searching attempts. This simplifies the discussion and leads to consider about $7.5\%$ of $N_T = 100 \times 10^9$ stars, that is $N_G = 7.5 \times 10^9$ stars and hence that  $R_S = N_G /N_T = 7.5 \times 10^{-2}$. 

As already stated, the remaining terms of the Drake equation are, but the last one, of probabilistic nature. The last term will be considered separately and will be suitably reinterpreted. 
In what concerns the terms $f_S$, $f_P$, $n_e$, $f_L$, $f_I$ and $f_T$, they correspond to conditional probabilities which are related to at least one previous condition, namely the one that appears immediately before in Drake's equation.   

We recall that the fundamental assumption relating two events that are conditioned is the Bayesian probability. If an event $A$ is conditioned by event $B$, then the probability of an event $A$ once $B$ occurs, $P(A/B)$, is given by:
\begin{equation}
P(A/B) = {P(B/A) P(A) \over P(B)}~,
\label{Bayes}
\end{equation}
where $P(B/A)$ is the probability of the event $B$ once $A$ occurs, while $P(A)$ and $P(B)$ are the probability of occurrence of events $A$ and $B$, respectively. 

At first glance, the fact that terms in Drake's equation are conditioned does not provide any obvious advantage. However, if we consider the string of terms, then an interesting simplification arise, namely:
\begin{equation}
 f_P n_e f_L f_I f_T  = {P(f_P/f_S) P(f_P/n_e) P(n_e/f_L) P(f_L/f_l) P(f_l/f_T) f_T  \over f_S}~,
\label{Bayesresult}
\end{equation}
where we have used the somewhat redundant notation of $P(f_X/f_Y)$ as the conditional probability of $X$ once $Y$ occurs.  The main point here is that as far as one can envisage: planets arise around stars, as hypothesised by Safranov \cite{Safranov}, once a stable disk is cleared from debris as corroborated by recent observable evidence \cite{RO}; in order to life to arise and to get developed into complex forms, planets are the most suitable niche for this development; advanced forms of life are the outcome of a complex evolution from early simple forms of life as the phylogenetic tree of all life species on Earth indicate. It is fairly plausible that this assumption remains valid elsewhere;  intelligent forms of life that give rise to civilisations, presumably, can only arise from complex forms of life, which, as recently point out, is greatly dependent on geological evolution that allows for continents, oceans and plate tectonics \cite{RStern}; willingness and capability to communicate is more likely to arise in the context of rather advanced civilisations. Thus, it is reasonable to conclude that as conditional probabilities are in the range $10^{-1} \le P(A/B) < 1$, then: 
\begin{equation}
 f_P n_e f_L f_I f_T \le O(10^{-4}) { f_T   \over f_S}~,
\label{result}
\end{equation}
and hence gathering the terms in a convenient way, we can say that if one excludes the $L$ term, Drake's equation gives rise to a probability, $P_D$,
\begin{equation}
 P_D =: R_S \times O(10^{-4})  { f_T   \over f_S}  = 7.5 \times 10^{-6} { f_T   \over f_S} ~.
\label{resultD}
\end{equation}

Therefore, in order to get an estimate for $P_D$ we have to analyse only the two remaining terms, $f_S$ and $f_T$. We shall address the estimate of these terms now and leave the discussion on $L$ for the next section. 

In order to estimate the number of stars that are surrounded by planets, we consider that the Milky Way has about 100 thousand millions stars and that the mechanism to form planetary disks is fairly general and driven by the capability of the star's gravity to attract interstellar matter which is well spread in the arms of the galaxy. It is thus quite feasible that $f_S$ is not too small, say $10^{-3}$ and thus: 
\begin{equation}
 P_D = 7.5 \times 10^{-3}  f_T   ~.
\label{resultD3}
\end{equation}
As for $f_T$, a number of inevitably speculative considerations are in order. Technological development takes place in the context of civilisations which understand that advances achieved through science and its applications are better suited to face the challenges of perpetuating their continuity within a specific planetary niche. It is somewhat consensual that sustainable technological development requires stability or a bearable amount of pressure for solving inner and/or extrinsic challenges. This evolution, likewise biological and sociological ones, are irregular and random, swift  and  abrupt at times, mediated by long periods of stagnation. These periods of immobility are often caused by famine, pandemics, wars and widespread damage to the environment due to the adoption of a predatory socio-economical model or through the use of destructive technologies. These are the very challenges that humankind faces right now. 

Anthropologists, economists, philosophers and sociologists have developed models to understand cultural, economical and social development. Under quite broad conditions, even though conjectural, the most accepted scenario supposes that groups of individuals start by hunting and gathering food till the limit of sustainability of a given region, and then  move to the harvesting of agricultural and animal products. Once these foundational events are achieved, the developed means of production ensure the multiplication of the population and a civilisational status is reached, so to admit a state, a urbanisation process, social stratification and a writing system. From these basic civilisational structures, societies  evolve to the harnessing of energy from the combustion of natural resources, such as coal, oil and gas to nuclear energy, and eventually to the storing and handling of information. IN the human context, the development of communication technology is needed to supply food and energy to a web of complex societies. The development of all these dimensions of harnessing took, on Earth, thousand of years and involved the effort of the whole humankind.  As is known, not all civilisations were able to reach significant progress in all the components of scientific and technological progress, most fundamentally because they were geographically isolated or because they have closed themselves from the world. In the latter case,  these civilisations assumed that they had nothing to profit from the exchange with other human groups. From these considerations, based on the fraction of civilisations, past and present, that in the human context thrived, one can advance the rough estimate, $f_T \simeq10^{-3}$, that is, about one in a thousand civilisations do evolve to become a technological civilisation.    

Of course, the above considerations concern specificities of the historical trajectories of the civilisations that thrived or perished under the context of our planet. However, on the other hand, instead of comparing the human civilisations among themselves, one could argue that the differences among putative civilisations in different planets are much wider and therefore it would be more suitable to consider, as a basis for comparison, the $8.2 \times 10^6$ different living species on Earth. This latter assumption would render an estimate of order $f_T \simeq 10^{-7}$. In what follows we shall consider the intermediate value, say $f_T \simeq 10^{-4}$, and hence:
\begin{equation}
 P_D \simeq 7.5 \times 10^{-7}  ~.
\label{resultD4}
\end{equation}
The uncertainties here can be of one or two orders of magnitude. In any case, given these considerations, in order to estimate the number of civilisations technologically capable of communication, requires an estimate for the $L$ parameter. This will be carried out in more detail in the next section.

\section{Estimating the time span to come across a cosmic civilisations based on some general assumptions about a systematic search}
\label{sec:iEstimateII}

Lets us now discuss the estimate of the parameter $L$. In the original Drake equation, the $L$ parameter was regarded as the time effort devoted to the task of searching for evidence on the existence of alien civilisations capable/willing to communicate. Even though this is a quite sensible choice, we think that it is advantageous to break this parameter into three pieces:  $f_E$ to represent the relative amount of energy, $E$, that a civilisation is willing to commit for detection, searching or communication with other cosmic civilisations; $f_A$, the resolution to be considered  in the task of searching a given cosmic region surrounding a G-type star; and, $f_S$, the impact that this effort  will have on the climatic equilibrium of the planet. In terms of Earth's climate crisis, it corresponds to the extent this effort will represent in terms of the  total heat generated to drive Earth away from the Holocene to the Anthropocene.  Explicitly, we propose rewriting $L $ in terms of three factors, each of them with specific time frames related to the nature of the component of the searching task they correspond to. Clearly, in broad terms, $f_E$ concerns limitations associated to the First Principle of Thermodynamics, while $f_S$ is related to the Second Principle of Thermodynamics.  

Thus, given that $L$ has units of time and the three components to be introduced can have each their own time dependence, then, we assume that
\begin{equation}
 L = [ f_E f_V f_S]^{1/3}~,
\label{L1}
\end{equation}
where
\begin{equation}
f_E = { E \over E_T} \Delta t_E~,
\label{L2}
\end{equation}
so that $E$ corresponds to the typical energy committed in the search and in data storing, handling and analysing,  $E_T$ to the total energy output of a civilisation and $\Delta t_E$ the time dedicated in the searching. The figures of merit for Earth overall transformation/generation of energy is $E_T \simeq 19.6 ~TW$ (2021 figure), and assuming that the commitment in searching for a cosmic civilisations involves, say the typical output of a nuclear power plant, namely $E \simeq 1~GW$, then for a time interval, 
$\Delta t_E$:
 \begin{equation}
f_E \simeq 5.0 \times 10^{-5} \Delta t_E ~.
\label{L3}
\end{equation}

In order to compute the factor $f_A$, we consider as a typical survey the observation of a region of the size of the solar system's planetary disk, that is, $7.7 \times 10^{-5}$ parsec around a G-type star at a distance of about $50$ parsecs so to safely avoid that the signal degrades down to noise. The beam resolution of a radio antenna is given by $\theta = 1.2 2~\lambda/D$, where $\lambda$ is the wavelength, typically between millimetre and  $10^2$ m, and $D$ the diameter of the antenna. Thus assuming that a survey around the star lasts a time $\Delta t_A$, we get: 
\begin{equation}
f_A \simeq \left( 1.5 \times 10^{-6} \over 1.22 \lambda/D\right)^2 \Delta t_A  \simeq 1.5 ~ \Delta t_A ~.  
\label{L4}
\end{equation}
Assuming $\lambda \simeq 10^{-3}$ m and $D \simeq 10^3$ m, which assumes already some mastery of radio interferometric science to  improve detection and searching capabilities. As is well known, interferometric techniques allow, through adding up baseline antennas, to achieve greater effective diameter and higher resolution. Furthermore, the rotation of the planet allows for the synthesis of imaging and setting telescopes into different configurations so to obtain higher resolution and to measure different spatial scales. Extending the interferometry capability into space and/or, 
for instance, to a moon, can enhance resolution considerably. It should be pointed out that recent SETI efforts involve already some interferometric capabilities. Here one could also consider the possibility that an alien civilisation  
detects our inadvertent radio emissions and sends back an aimed signal, which would allow for existing radio telescopes to detect these aimed signal even from very far away. Of course, it is hard to quantify this possibility, but it seems sensible to consider its likelihood to be rather low so that the uncertainty it brings in our estimates not very relevant.  

Finally, in order to estimate the factor $f_S$ we consider the Second Principle of Thermodynamics and assume typically that the efficiency $\eta$, of the energy convertors/generators are of $ 40 \%$, meaning that $60 \%$ of the total energy output considered above, $E_T$ is released to the environment as heat. Furthermore, assuming that this heat raises the planet's average temperature, $T_M$, by $\Delta T$, hence using the black-body approximation and Stefan's law, $P=  \sigma A T^4$, where $\sigma = 5.670 \times 10^{-8}~W m^{-2} K^{-4}$ is the Stefan-Boltzmann constant, $A$ the area of the planet and $T$ its absolute temperature, then: 
\begin{equation}
f_S = {0.6 E_T   \over 4 \pi R_T^2 \sigma \left[ (T_M+ \Delta T)^4 - T_M^4\right]} \Delta t_S~. 
\label{L5}     
\end{equation}
Thus, for typical Earth's values, $T_M \simeq 288 K$, and $\Delta T = 1.5~K$ as the ,maximal allowed value, one obtains:
\begin{equation}
f_S = 2.83 \times 10^{-3} \Delta t_S~. 
\label{L6}
\end{equation}

It is just natural that  $\Delta t_E = \Delta t_A =\Delta t_S$, and  using the above results, Eqs. (\ref{resultD4}),  (\ref{L3}), (\ref{L4}) and (\ref{L6}),  considering that the search can go beyond $G$-type stars in our galaxy, from which one can get a factor 2 or so, hence:
\begin{equation}
N= 1.4 \times 10^{-8} \Delta t_A~. 
\label{L7}
\end{equation}

Therefore, requiring that $N \ge1$ and assuming about 10 frequencies about the $21$ cm bandwidth, then $\Delta t_A  \ge 22$ years, meaning that a search of a hundred stars might last a few millennia . Thus, it seems reasonable to conclude that a successful  searching task would require means beyond the capability of a Kardashev Type-I civilisation.  Notice that for a more conservative choice of factors, would reinforce the conclusion that a Kardashev Type-I civilisation could hardly expect for success in the search of advanced forms of life.  It should be pointed out that for a Kardashev Type-II civilisation it should be easy to improve on the $f_E$ factor, but it will inevitably face problems in what concerns factor $f_S$, that is to say it would find difficult to handle the climate crisis it engenders.

\section{Discussion and Conclusions}
\label{sec:iConclusions}

The foundational basis of a science like astrobiology is somewhat hard to setup. The continuous transformation of Earth conditions make it impossible to establish the fundamental conditions that gave origin to life and the uncountable contingent changes that shaped biological evolution. These difficulties do not allow even excluding for sure the hypothesis that life has come to Earth from elsewhere, even though the apparent widespread sterile conditions of the Solar System suggest that  any process of seeding, casual or deliberate, would leave footsteps elsewhere besides Earth. Furthermore, the highly structured biochemical combinations that characterise life suggest that it might more likely arise within an ever changing environment typical of an ever evolving planet, rather than under the harsh and hostile space conditions.  Of course, all these considerations might be faulty as they rely on assumptions based on an extremely limited amount of hard theoretical and empirical evidence. 

Of course, under these clearly shaky grounds, it is somewhat hopeless to achieve reliable quantitive estimates even about the most basic issues. This is even more questionable  when further considering downstream matters like the ones encapsulated by Drake's equation. No doubt that this equation is a quite brave attempt to setup the guidelines for a scientific programme and it has the merit to spring from the observational specificities of radioastronomy, an already mature and solidly settled branch of astronomy.  It is worth mentioning that radioastronomy is at the verge of a qualitative and quantitative upgrading with the world wide SKA project (see, for instance, Ref.  \cite{SKA} for a broad discussion). and it can greatly boost attempts for searching signs of the existence of extraterrestrial intelligence as discussed above. Furthermore, the possibility of structuring different observational strategies, such as SETI, Communication with Extraterrestrial Intelligence (CETI) or Messaging to Extra-Terrestrial Intelligence (METI or Active SETI) shows that there is already a scientific culture and even infrastructure to carry out observational campaigns based on some minimally consensual scientific standards.   

However, it is clear that Drake's equation encompasses too many issues, including of anthropological, biological, historical and socio-economical nature, about which our knowledge on Earth, although deep, is predominantly qualitative. This makes a quantitative analysis, even in bulk as proposed by Drake's equation, somewhat difficult to support and even questionable in its usefulness for any type of accumulative knowledge that goes beyond  the basic, ``to be"  or "not to be" assessment. 

In this work, we have attempted to introduce more structure into Drak'e's equation. This was done through two different approaches. Firstly, we have identified a conditional structure within the probabilistic terms that ranged from fundamental planetary formation to biology, from civilisational and technological matters and shown that through the conditional Bayesian probability one can obtain some simplification in estimating the various terms in question.  Secondly, we have argued that the time dedicated for the search of extraterrestrial civilisations, can be densified by introducing the three most basic levels of observational analysis:  the energetic involvement in the task; the search resolution factor for a systematic study of a  patch of sky around a G-type star; and finally, the impact that the searching task has on the climate crisis on a Earth-like planet. 

Our assumptions are rather conservative as they assume that an energy supply of a nuclear plant with about $40 \%$ of efficiency is used in the searching task, that searches around a G-type star involve realistic resolution and some interferometric capabilities so that the task does not lead to significant climate change. The modesty of the task and the amount of time it demands, about $20$ years, for searching in 10 frequencies, suggest that  a Kardashian type-I civilisation is at the verge of being capable to carry out an effective research plan that has some chances of success . We stress that the climate crisis issue must be necessarily included in any long term scientific project, an astrobiology one being no exception. 

Of course, it is not at all unlikely that in a subject like astrobiology, a series of unexpected breakthroughs are still in stock, from the basic biochemistry level to the way how intelligent forms of life can get organised. This means that considerations based on sensible astrophysical arguments and on Drake's equation, in particular, might be too naive to capture the complexity of the problem Nevertheless, we believe that till the discovery of ground breaking new facts, broad quantitative estimates like the ones provided by Drake's equation and its variations might be still useful to guide our gaze through the clouds of these fascinating matters.


\newpage
\vspace{0.5cm}

{\bf Acknowledgments~~}

\noindent
The author would like to thank Adam Frank and Clovis de Matos for interesting discussions and suggestions. 

\vspace{0.3cm}


\bibliographystyle{unsrtnat}

\end{document}